\documentclass[preprint,showpacs,preprintnumbers,amsmath,amssymb,natbib]{revtex4}

\usepackage{graphicx}
\usepackage{epsfig}		
\usepackage{dcolumn}
\usepackage{bm}

\def\0{\mbox{\tiny $0$}}
\def\1{\mbox{\tiny $1$}}
\def\2{\mbox{\tiny $2$}}
\def\3{\mbox{\tiny $3$}}
\def\4{\mbox{\tiny $4$}}
\def\5{\mbox{\tiny $5$}}
\def\6{\mbox{\tiny $6$}}
\def\7{\mbox{\tiny $7$}}
\def\8{\mbox{\tiny $8$}}
\def\9{\mbox{\tiny $9$}}

\def\f14{\mbox{\tiny $\frac{1}{4}$}}

\DeclareMathOperator{\Tr}{\mbox{Tr}}

\begin{document}

\title{Entanglement of Dirac bi-spinor states driven by Poincar\'e classes of $\mbox{SU}(2) \otimes \mbox{SU}(2)$ coupling potentials}

\author{Victor A. S. V. Bittencourt}
\email{vbittencourt@df.ufscar.br}
\author{Alex E. Bernardini}
\email{alexeb@ufscar.br}
\affiliation{Departamento de F\'{\i}sica, Universidade Federal de S\~ao Carlos, PO Box 676, 13565-905, S\~ao Carlos, SP, Brasil}

\date{\today}

\begin{abstract}
A generalized description of entanglement and quantum correlation properties constraining internal degrees of freedom of Dirac(-like) structures driven by arbitrary Poincar\'e classes of external field potentials is proposed.
The role of (pseudo)scalar, (pseudo)vector and tensor interactions in producing/destroying intrinsic quantum correlations for $\mbox{SU}(2) \otimes \mbox{SU}(2)$ bi-spinor structures is discussed in terms of generic coupling constants.
By using a suitable \textit{ansatz} to obtain the Dirac Hamiltonian eigenspinor structure of time-independent solutions of the associated Liouville equation, the quantum entanglement, via concurrence, and quantum correlations, via geometric discord, are computed for several combinations of well-defined Poincar\'e classes of Dirac potentials.
Besides its inherent formal structure, our results set up a framework which can be enlarged as to include localization effects and to map quantum correlation effects into Dirac-like systems which describe low-energy excitations of graphene and trapped ions.
\end{abstract}


\date{\today}
\maketitle

\section{Introduction}

The map of controllable physical systems onto the Dirac equation formal structure has been in the core of recent investigations which search for reproducing quantum relativistic effects, such as the \textit{zitterbewegung} effect and the Klein tunneling/paradox, on table-top experiments \cite{n001,n002,n003,n004,n005,n006}. For instance, the low energy excitations of bilayer graphene has been effectively described by a Dirac-like equation \cite{n002}, in the same route that the engendering of Jaynes-Cummings Hamiltonian has allowed to simulate Dirac dynamics with a single trapped ion \cite{n001}.
On the front of analogies with the high energy physics, examples of such connection are the black-hole properties in Bose-Einstein condensates \cite{n007}, the simulation of Unruh effect in trapped ions \cite{n008} and the \textit{trembling} motion and the Klein's paradox for massive fermions in 2D systems \cite{n009}.

Dirac equation was first presented as a relativistic description of quantum mechanics \cite{n010} as to preserve the quantum probabilities axioms.
In the scope of Coulombian interactions, the fine structure of the Hydrogen atom was derived and the first prediction of existence of antiparticles \cite{n011} has been formally presented.
On the other hand, given that Dirac solutions support a $\mbox{SU}(2)\otimes \mbox{SU}(2)$ group structure involving two internal degrees of freedom, the \textit{spin} and the \textit{intrinsic parity}, the free particle solutions of the Dirac equation may exhibit \textit{spin-parity} entanglement \cite{n012,n013}.
Recent results \cite{n012,n013} indeed establishes that the $\mbox{SU}(2) \otimes \mbox{SU}(2)$ representation of Dirac bi-spinors are assigned to a Hamiltonian dynamics written in terms of the tensor (direct) product of two-qubit operators, ${H}_{D}={\sigma}_{x}^{\left( 1\right) }\otimes \left(
\vec{p}\cdot \vec{{\sigma}}^{\left( 2\right) }\right) +m \,
{\sigma }_{z}^{\left( 1\right)}\otimes {I}^{(2)}_{2}$,
through which the free particle Dirac equation solutions are given in terms of {\em spin-parity} entangled states.
In such a context, the interface of relativistic quantum mechanics and quantum information provides a robust framework to classify and quantify the information content of Dirac bi-spinors and, in particular, the influence of Poincar\'e transformations on their correlational content \cite{n014}.

The inclusion of quantum potentials driven by external fields in the Dirac Hamiltonian is relevant for the study of a wide variety of physical systems \cite{n010}, from the Hydrogen atom \cite{n011} to hadronic \cite{n015,n016} and nuclear physics \cite{nucl001,nucl002,nucl003} models.
Such external fields are included into the Dirac equation through the addition of potential matrices to the free Hamiltonian, and the possible terms of such matrices are derived by considering the invariance of the resulting Dirac equation under Poincar\'e transformations \cite{n010}.
In the scope of Dirac-like systems described by the Dirac equation with external potentials, one can notice, for instance, that the Jaynes-Cummings Hamiltonian can mimic such external fields \cite{n017}, or even lattice imperfections in graphene can be described by external field effects on the free Dirac dynamics \cite{n018}.

External fields also change the entanglement content of Dirac bi-spinors. In a previous issue, one has investigated how a barrier scattering, implemented by a static electric potential via minimal coupling in the Dirac equation, can either create or destroy ({\em spin-parity}) entanglement for free particle states \cite{n019}.
This has inaugurated a new context in which the entanglement content of bi-spinors can be mapped into Dirac-like systems, providing a framework to compute, for example, \textit{spin-spin} entanglement of nonrelativistic systems, namely, for electron-hole pairs in graphene, and low-energy excitations of trapped ions.

In this paper one shall identify a systematic approach for deriving \textit{spin-parity} entanglement properties of Dirac equation bi-spinor solutions driven by several Poincar\'e classes of Dirac potentials. An \textit{ansatz} for obtaining stationary solutions of the Liouville equation related to the Dirac Hamiltonian is proposed and applied for identifying and quantifying entanglement and correlation properties of the corresponding quantum states.
Explicit analytic results are obtained for external fields organized into four particular categories: the pseudoscalar potential, the pseudoscalar plus tensor potentials, the pseudoscalar plus pseudovector potentials and the pseudoscalar plus pseudovector and tensor potentials. For all of them, vector potential is implicitly included given that it transforms as the kinetic term under Poincar\'e symmetries.
Entanglement is quantified by the quantum concurrence as function of generalized phenomenological parameters such as, for instance, the relative angle between interacting fields, or between fields and particles momentum.
For the above-mentioned first case involving only the pseudoscalar potential, the geometric discord is also used to quantify quantum correlations.

The manuscript is organized as follows.
In Sec. II, the background formalism for identifying the entanglement/separability information content of Dirac bi-spinors is established, and its correspondence with the classes of potential matrices following the prescriptions of \cite{n010} is presented.
In Sec. III, the discussion about entanglement and quantum correlations is brought up in order to set up the corresponding expressions for the quantum concurrence and for the geometric discord that are derived and interpreted as the quantifiers of the informational content of interacting Dirac bi-spinor states. The \textit{ansatz} for the density matrix is introduced in Sec. IV, where the formalism which allows for computing the entanglement of the above-mentioned external potential configurations is implemented. Sec. V contains our final conclusions, where remarks and urging perspectives related to the physics of graphene and trapped ions are discussed.

\section{Dirac equation and Poincar\'e classes of interacting potentials}

Dirac equation was proposed as a relativistic invariant wave equation that possess a non negative conserved probability density. In its Hamiltonian form, the Dirac equation reads \cite{n010}
\begin{equation}
\label{E01}
i \frac{\partial}{\partial t} \psi (\vec{x}, t) = H_0 \psi(\vec{x},t) = (\, \vec{p} \cdot \vec{\alpha} + m \beta \,) \psi(\vec{x},t),
\end{equation}
where $\vec{p}$ is the momentum, $m$ is the particle mass and one has considered natural units with $\hbar = c = 1$. $\vec{\alpha}$ and $\beta$ are anticommuting quantities represented by $n \times n$ matrices which must satisfy
\begin{eqnarray}
\alpha_i \alpha_j + \alpha_j \alpha_i &=& 2 \delta_{i,j} I_4, \nonumber \\
\alpha_i \beta + \beta \alpha_i &=& 0, \nonumber \\
\beta^2 = I_4,
\end{eqnarray}
in order to impose the free particle dispersion relation $E^2 = p^2 + m^2$ over (\ref{E01}), where throughout the paper, $p = \vert\vec{p}\vert$, with definition extended to any vector quantity, $v$.

For the $3+1$ dimensional space-time representation, $n$ must be even and the minimum dimensionality of $\vec{\alpha}$ and $\beta$ is $4$. These matrices have different representations interconnected by unitary transformations, and hereafter one may consider one of them as set up by
\begin{eqnarray}
\vec{\alpha} = \sigma_x \otimes \vec{\sigma} &=&  \left[ \begin{array}{rr} 0 & \vec{\sigma} \\ \vec{\sigma} & 0 \end{array}\right], \nonumber \\
\beta = \sigma_z \otimes I_2 &=& \left[ \begin{array}{rr} I_2 & 0 \\ 0 & - I_2 \end{array} \right],
\end{eqnarray}
where $\vec{\sigma}$ are the Pauli matrices.  Eq.~(\ref{E01}) can be written in a covariant form in terms of the Dirac matrices $\gamma_\mu = (\beta, \beta \vec{\alpha})$ as \cite{n010}
\begin{equation}
(\gamma_\mu p^\mu - mI_4) \psi(x) = 0,
\end{equation}
with $p^\mu$ describing the four momentum operator $ \sim i \partial_\mu$ and $x = (t, -\vec{x})$. The Dirac matrices satisfy the anticommutation relation \begin{equation}
\label{A01}
\{ \gamma_\mu , \gamma_\nu \} = 2 \eta_{\mu \nu},
\end{equation}
where $\eta_{\mu \nu}$ is the Minkowski metric tensor.

Any product of $\gamma_\mu$ matrices can be reduced to linear and bi-linear compositions through the relation (\ref{A01}), which is used to eliminate pairs of identical matrices.
It restricts the number of irreducible products of Dirac matrices to $16$ traceless elements, $\{\Gamma_i;\,i =0,\ldots 15 \}$, defined as
\begin{eqnarray}
\Gamma_0 &\equiv& I_4, \nonumber\\
\{\Gamma_1, \, \dots, \, \Gamma_4\} &\equiv& \gamma_\mu, \nonumber \\
\Gamma_5 &\equiv& \gamma_5 = i \gamma_0 \gamma_1 \gamma_2 \gamma_3, \nonumber \\
\{\Gamma_6, \, \dots, \, \Gamma_9\} &\equiv& \gamma_5 \gamma_\mu, \nonumber \\
\{\Gamma_{10}, \, \dots, \, \Gamma_{15}\} &\equiv& \frac{i}{2} \{ \gamma_\mu , \gamma_\nu \}.
\end{eqnarray}
Besides their linearly independency and their mutual orthogonality, these matrices span the vector space of $4\times4$ matrices, i.e., any $4\times4$ matrix $X$ can be written as a linear combination $X = \sum_i x_i \Gamma_i$.
Therefore, the introduction of external potentials in the Dirac theory is performed through the addition of a $4 \times 4$ Hermitian matrix $V$ to the free Dirac Hamiltonian $H_0$,
\begin{equation}
i \frac{\partial}{\partial t} \psi (\vec{x}, t) = H \psi(\vec{x},t)= (H_0 + V) \psi(\vec{x},t).
\end{equation}
Within this framework $V = V(\vec{x},t)$ includes six terms that are classified according to their transformation properties under Poincar\'e transformations \cite{n010}. In fact, the covariant form of the above equation
\begin{equation}
\label{E02}
(i \gamma^\mu \partial_\mu - mI_4 - U (x))\psi(x) = 0,
\end{equation}
where $U (x) = \gamma_0 V (\vec{x}, t)$ is the covariant potential, transforms under a Poincar\'e transformation $(a, L) \in \mathcal{L}^\uparrow$ as
\begin{equation}
(i \gamma^\mu \partial_\mu - mI_4 - U^\prime (x))\, L \, \psi(\Lambda_L ^{-1}(x - a)) = 0.
\end{equation}
To (\ref{E02}) be invariant, $U$ must follow the transformation law,
\begin{equation}
\label{E09}
U^\prime (\vec{x}, t) = L \, U(\Lambda_L ^{-1}(x - a)) \, L^{-1},
\end{equation}
and given that $\gamma_0 \, L \, \gamma_0 = {L^*}^{-1}$, the potential $V$ must transform as
\begin{equation}
\label{E03}
V^\prime (x) = {L^*}^{-1} \, V(\Lambda_L ^{-1}(x - a)) L^{-1}.
\end{equation}
Moreover, if a wave function $\psi$ satisfies the Dirac equation (in its Hamiltonian form) with the potential $V$, then the Poincar\'e transformed wave function $L \, \psi(\Lambda_L ^{-1}(x - a))$ is a solution of the Dirac equation with the potential given by (\ref{E03}) \cite{n010}. By exploring transformation properties one can derive six classes of covariant potentials, $U$.

For a real valued function $\phi_S(x)$ that transforms under a Poincar\'e symmetries as $\phi_S (x) \rightarrow \phi_S (\Lambda^{-1}(x-a))$, a {\it scalar potential} is defined by
\begin{equation}
U_S = \phi_S (x) \longrightarrow V_S = \gamma_0 \,\phi_S (x),
\end{equation}
and can be considered, for instance, to introduce the mass term into the Dirac Hamiltonian.

The real function $\mu(x)$ defines the {\it pseudoscalar potential} as
\begin{equation}
U_{PS} = i \gamma_5 \,\mu(x) \longrightarrow V_{PS} = i \gamma_0 \gamma_5\, \mu (x).
\end{equation}
Although $\mu(x)$ behaves as a scalar invariant under Poincar\'e transformations, to set the Dirac equation invariant under spatial reflection, $P$, one has to assume that $\mu(x) = - \mu (P \,x)$.

The {\it vector potential} is introduced by
\begin{equation}
U_V = \gamma_\mu A^\mu (x) \longrightarrow V_V =  A^0(x) - \vec{\alpha} \cdot \vec{A}(x),
\end{equation}
where $A^\mu (x)$ transforms as a vector, i.e. $A^\mu(x) \rightarrow (A^\nu)^\prime (x) = \Lambda^\nu _{\, \mu} A^\mu (\Lambda^{-1}(x-a))$.
The vector interaction term is included into the Hamiltonian via minimal coupling, i. e. by replacing $H$ by $\tilde{H} = H - A^0\,I_4$ and $\vec{p}$ by $\vec{\mathcal{P}} = \vec{p} - \vec{A}$.
As an example, electromagnetic interactions driven by electric and magnetic fields correspond to $\vec{E} = - \vec{\nabla} A^{0}(x) - \partial_t \vec{A} (x)$ and $\vec{B} = \vec{\nabla} \times \vec{A}$.

The {\it pseudovector potential} is defined by means of four real functions $W_\mu =(q(x), \,\vec{W}(x))$ as
\begin{equation}
U_{PV} = \gamma_5 \gamma_\mu W^\mu (x) \longrightarrow V_{PV}(x)=- \gamma_5 \,q(x) + \gamma_5 \,\vec{\alpha} \cdot \vec{W}(x),
\end{equation}
which transform as a vector field under Poincar\'e symmetries.
Otherwise if, under spatial reflections, it transforms as $W_\mu(x) = - P\, W_\mu (P\,x)$, then Dirac equation is invariant, and thus $W^\mu$ transforms as a pseudovector.

For an antisymmetric field transforming as $C^{\mu \nu} (x) \rightarrow\Lambda_{\varrho} ^{\, \mu} \Lambda_{\tau} ^{\, \nu} C^{\varrho \tau} (\Lambda^{-1}(x - a))$, that is a tensor field, the {\it tensor potential}
\begin{equation}
U_{T} = \frac{\,\kappa_a\,}{2} \sigma^{\mu \nu} \,C_{\mu \nu} (x),
\end{equation}
is invariant under Poincar\'e transformations.
One can identify $C_{\mu \nu}$ as the electromagnetic tensor $F_{\mu \nu} = \partial_\mu A_ \nu - \partial_\nu A_\mu$, which can, for instance, describe the interaction between a Dirac field and an electromagnetic field \cite{n020} through a magnetic moment coupling, $\,\kappa_a\,$.
It also has been applied to compute spin-orbit interactions in atomic nuclei \cite{nucl001,nucl002,nucl003}.
The tensor components of the Hermitian potential can be written in terms of the electric and magnetic fields as
\begin{equation}
V_{T} = \,\kappa_a\,( i \vec{\gamma} \cdot \vec{E}(x) + \gamma_5 \vec{\gamma} \cdot \vec{B}(x) ).
\end{equation}

Again, by means of an antisymmetric (tensor) field correspondence, $D_{\mu \nu}$, the {\it pseudotensor potential} is defined by
\begin{equation}
U_{PT} = - i \frac{\chi_a}{2} \gamma_5 \sigma^{\mu \nu} D_{\mu \nu} (x).
\end{equation}
Similar to pseudoscalar and pseudovector interactions, the invariance of Dirac equation under spatial reflections constrains $D_{\mu \nu}$ to transform as a pseudotensor. If $D_{\mu \nu}(x)$ is taken as the electromagnetic tensor, the Hermitian potential is written as
\begin{equation}
V_{PT} = \chi_a (i \vec{\gamma} \cdot \vec{B}(x) - \gamma_5 \vec{\gamma} \cdot \vec{E}(x)),
\end{equation}
and accounts for the interaction of the Dirac field with the electromagnetic field coupled by an electric moment, $\chi_a$. However, for this case, the Dirac equation is not invariant under spatial reflections once that the electromagnetic tensor does not invert its sign under spatial reflections.

Table \ref{Tab:01} summarizes the potential matrices in their covariant $U(x)$, Hermitian $V(x) = \gamma_0 U(x)$ and $\mbox{SU}(2) \otimes \mbox{SU}(2)$ decomposition formats, in particular, for tensor and pseudotensor potentials given in terms of the electromagnetic tensor fields, $F_{\mu \nu} (x)$.

\begin{table}[h]
\resizebox{16 cm}{!}{
\begin{tabular}{| c| c| c| c |}
\hline
\hspace{.5 cm} Potential \hspace{.5 cm}\hspace{.5 cm} &\hspace{.5 cm}  $U(x)$ \hspace{.5 cm} &\hspace{.5 cm} $V(x)$\hspace{.5 cm} & \hspace{.5 cm} $\mbox{SU}(2)\otimes \mbox{SU}(2)$ form \hspace{.5 cm} \\ \hline
Scalar & $\phi_S (x)$ & $\gamma_0 \phi_S (x)$ &$[\sigma_z ^{(1)} \otimes I^{(2)}] \phi_S(x)$ \\ \hline
Pseudoscalar & $i \gamma_5 \mu(x)$ & $i \gamma_0 \gamma_5 \mu(x)$ & $-[\sigma_y ^{(1)} \otimes I^{(2)}] \mu(x) $\\ \hline
Vector &$ \gamma_\mu A^\mu (x) $ &$ A_0(x) I - \vec{\alpha} \cdot \vec{A}(x) $ & $ [I^{(1)} \otimes I^{(2)}] A_0(x) - [\sigma_x^{(1)} \otimes \vec{\sigma}^{(2)}] \cdot \vec{A}(x) $\\ \hline
Pseudovector &$ \gamma_5 \gamma_\mu W^\mu(x) $&$ - \gamma_5 q(x) + \gamma_5 \vec{\alpha}\cdot\vec{W}(x)$ & $-(\sigma_x^{(1)} \otimes I^{(2)}) q(x) + [I^{(1)} \otimes \vec{\sigma}^{(2)}]\cdot \vec{W(x)}$ \\ \hline
Tensor & $\frac{\,\kappa_a\,}{2} \sigma^{\mu \nu} F_{\mu \nu}(x)$ & $\, \,\kappa_a\,( i \vec{\gamma} \cdot \vec{E}(x) + \gamma_5 \vec{\gamma} \cdot \vec{B}(x) )\,$ & $\, - \,\kappa_a\, (\sigma_y ^{(1)}\otimes \vec{\sigma}^{(2)}) \cdot \vec{E}(x) -  \,\kappa_a\, (\sigma_z^{(1)} \otimes \vec{\sigma}^{(2)})\cdot \vec{B}(x)\, $ \\ \hline
Pseudotensor & $- i \frac{\chi_a}{2} \gamma_5 \sigma^{\mu \nu} F_{\mu \nu} (x)$ & $\, \chi_a (i \vec{\gamma} \cdot \vec{B}(x) - \gamma_5 \vec{\gamma} \cdot \vec{E}(x))\, $ & $\,-\chi_a (\sigma_y^{(1)} \otimes \vec{\sigma}^{(2)}) \cdot \vec{B}(x) +  \chi_a (\sigma_z^{(1)}\otimes \vec{\sigma}^{(2)})\cdot \vec{E}(x)\,$ \\ \hline
\end{tabular}
}
\caption{Possible potential matrices of Dirac Hamiltonian.}
\label{Tab:01}
\end{table}

The full Hamiltonian with all the interactions from Table \ref{Tab:01} reads
\begin{eqnarray}
\label{E04T}
H  &=& A^0(x)\,I_4+ \gamma_0( m + \phi_S (x) ) + \vec{\alpha} \cdot \vec{\mathcal{P}} + i \gamma_0 \gamma_5 \mu(x) - \gamma_5 q(x) + \gamma_5 \vec{\alpha}\cdot\vec{W}(x) \nonumber \\
&+& i \vec{\gamma} \cdot [ \chi_a \vec{B}(x) + \kappa_a\, \vec{E}(x)  \,] + \gamma_5 \vec{\gamma}\cdot[\kappa_a\, \vec{B}(x)  - \chi_a \vec{E}(x) \,].
\end{eqnarray}

\section{Bispinors and entanglement}

The $\mbox{SU}(2)\otimes \mbox{SU}(2)$ representation of Dirac bi-spinors is generated by the free Hamiltonian given in terms of two-qubit operators, ${H}_{D}={\sigma}_{x}^{\left( 1\right) }\otimes \left(
\vec{p}\cdot \vec{{\sigma}}^{\left( 2\right) }\right) +m \,
{\sigma }_{z}^{\left( 1\right)}\otimes {I}^{(2)}_{2}$ \cite{n012,n013} for whose the solutions, in momentum space, are written in terms of a sum of direct products describing \textit{spin-parity} entangled states
\begin{eqnarray}
\left\vert \Psi ^{s}(\vec{p},\,t)\right\rangle
&=&e^{i(-1)^{s}\,E_{p}\,t}\left\vert \psi ^{s}(\vec{p})\right\rangle
= e^{i(-1)^{s}\,E_{p}\,t}N_{s}\left( p\right)  \notag
\\
&\times& \left[ \left\vert
+\right\rangle _{1}\otimes \left\vert u(\vec{p})\right\rangle _{2}+\left(
\frac{p}{E_{p}+(-1)^{s+1}m}\right) |-\rangle _{1}\,\otimes \left( \hat{p}
\cdot \vec{{\sigma }}^{\left( 2\right) }\left\vert u(\vec{p}
)\right\rangle _{2}\right) \right],
\end{eqnarray}
where $s = 0$ and $1$ stands respectively for particle and antiparticle (intrinsic parity) associated frequency solutions and the spinor $u_s (p)$ is coupled to the spin states and it is related to the spatial motion of the respective solution.
The above representation sets up the interface between relativistic quantum mechanics and quantum information theory, where the discrete degrees of freedom of Dirac bi-spinors are associated to a system $\mathcal{S}$ composed of two subsystems, $\mathcal{S}_1$ (\textit{spin} system) and $\mathcal{S}_2$ (\textit{intrinsic parity} system), circumvented by a composite Hilbert space $\mathcal{H} = \mathcal{H}_1 \otimes \mathcal{H}_2$ with $\mbox{dim}\, \mathcal{H}_1 = \mbox{dim}\, \mathcal{H}_2 = 2$. The corresponding bi-partite states are called two-qubit states, and from this point, our discussion shall be concerned with these types of states.

Composite quantum systems, as those prescribed by Dirac bi-spinors, can exhibit quantum correlations like entanglement.
The entanglement is a consequence of the superposition principle and it is defined through the concept of separability. A bi-partite state is separable if it is possible to write its density operator $\varrho$ as \cite{n021}
\begin{equation}
\varrho = \sum_{i} w_i \, \sigma_i ^{(1)} \otimes \tau_i ^{(2)},
\end{equation}
where $\sigma_i ^{(1)} \in \mathcal{H}_1$, $\tau_i ^{(2)} \in \mathcal{H}_2$ and $\sum_i w_i = 1$. For pure systems, for whose density operators are of the form $\vert \psi \rangle \langle \psi \vert$, the separability definition can be stated for the state vector as $\vert \psi \rangle = \vert \psi_1 \rangle \otimes \vert \psi_2 \rangle$. If a state is not separable then it is entangled.

Entanglement is quantified through different methods, depending on the context and on the specific properties that are considered \cite{n022}. For pure states, the Schmidt decomposition theorem guarantees that the reduced density operators $\varrho_{1 \, (2)} = \Tr_{2 (1)}[\varrho]$ have the same eigenvalues and if the state is entangled then either $\varrho_{1(2)}$ are mixed states. In this case the entanglement entropy $E_{vN}[\varrho]$ is defined in terms of the von Neumann entropy $S[\varrho]$ of their subsystems as \cite{n021},
\begin{equation}
E_{vN}[\varrho] = S[\varrho_2] = - \Tr_2[\varrho_2 \log_2 \varrho_2] = S[\varrho_1] = - \Tr_1[\varrho_1 \log_2 \varrho_1],
\end{equation}
is a quantifier of the entanglement in $\varrho$. However, if the state is mixed, there is no guarantee that a mixed subsystem is associated to an entangled state. Entanglement quantifiers for mixed states can be built through the convex-roof extension of pure state entanglement quantifiers \cite{n023}. For instance, the entanglement of formation \cite{n024} is the convex-roof extension of the entanglement entropy, the average of the pure-state entanglement,
\begin{equation}
E_{EoF} [\varrho] = \mbox{min}_{\varrho_k} \displaystyle \sum_k p_k E_{vN}[\varrho_k],
\end{equation}
minimized over all decompositions of the mixed state $\varrho$ on pure states, $\varrho_k$. For two-qubit states, the entanglement of formation is given by
\begin{eqnarray}
E_{EoF} [\varrho] &=& \mathcal{E}\left[ \frac{1 - \sqrt{1 - (C [\varrho])^2}}{2}\right], \nonumber \\
\mathcal{E}[x] &=& - x \log_2 x - (1-x)\log_2 (1-x).
\end{eqnarray}
In this case entanglement of formation is a function of the concurrence $C[\varrho]$ defined by \cite{n024}
\begin{equation}
\mathcal{C}[\varrho] = \mbox{max}\{ \lambda_1 - \lambda_2 - \lambda_3 - \lambda_4 \, , \,0 \},
\end{equation}
where $\lambda_1 > \lambda_2 > \lambda_3 > \lambda_4$ are the eigenvalues of the operator $\sqrt{\, \sqrt{\varrho} \, (\sigma_y \otimes \sigma_y) \varrho^\ast (\sigma_y \otimes \sigma_y) \, \sqrt{\varrho}\, }$. Although its definition is primarily related to the entanglement of formation, concurrence is by itself an entanglement quantifier.

A generic two-qubit system can be written in the form of
\begin{equation}
\label{D04}
\varrho = \frac{1}{4} \left[ I + (\vec{\sigma}^{(1)} \otimes I^{(2)}) \cdot \vec{a}_1 + (I^{(1)} \otimes \vec{\sigma}^{(2)}) \cdot \vec{a}_2  + \displaystyle \sum_{i,j = 1}^3 t_{ij} (\sigma_i^{(1)} \otimes \sigma_j^{(2)}) \right],
\end{equation}
where $\sigma_i$ are the Pauli matrices, $[T]_{ij} = t_{ij}$ is the correlation matrix and $\vec{a}_{1 \, (2)}$ are the Bloch vectors of the corresponding subsystem. For pure states $a_1 ^2 = a_2 ^2$, the concurrence is given in terms of the Bloch vectors as
\begin{equation}
\mathcal{C}[\varrho] = \sqrt{1 - a_{1}^2} = \sqrt{1 - a_{2}^2}.
\end{equation}

Separable mixed states can exhibit quantum correlations other than entanglement \cite{n025} and their characterization is still a partially open problem.
Quantum discord, for instance, is a measure of nonclassical correlations and it is defined as the difference of the total correlation between two subsystems before and after a perfect local measurement process in one of them \cite{n026}. The quantum discord related to measurements in the subsystem $1$ is computed through the complete set of projectors in that subsystem, $\{\Pi_k ^{(1)}\}$, as
\begin{equation}
\mathcal{D} [2\vert \vert1] = S[\varrho_2] - S[\varrho] + \mbox{max}_{\{ \Pi_k ^{(1)}\}} \displaystyle \sum_k p_k S[\varrho_{2 \vert \vert k}],
\end{equation}
where $\varrho_{2 \vert \vert k}$ is the state of the subsystem $2$ after the measurement $\Pi_k$ be performed on $1$,
\begin{eqnarray}
\varrho_{2 \vert \vert k} &=& \frac{\Tr_1 [\Pi_k \varrho \Pi_k]}{p_k}, \nonumber \\
p_k &=& \Tr[\Pi_k \varrho].
\end{eqnarray}
Since its definition involves an optimization process, it is impossible to derive an explicit formula for the quantum discord for a generalized state.

The study of the geometry of quantum correlations in the Hilbert space of states can also be adopted as another point of view over quantum correlations. In this context, the geometric discord $D[\varrho]$ is defined as the minimum distance between the state $\varrho$ and the set of states with zero quantum discord \cite{n027}. It has been demonstrated that $D[\varrho] = 0$ is a necessary and sufficient condition for vanishing quantum discord. For two-qubit states (\ref{D04}) geometric discord is analytically evaluated as
\begin{equation}
D_{1 \, (2)}[\varrho] = \frac{1}{4} \left( \, a_{1 \, (2)}  ^2 \, + \vert \vert \, T \, \vert \vert ^2 - k_{max} \right),
\label{disc}
\end{equation}
where $\vert \vert \, T \, \vert \vert ^2 = \mbox{Tr}[\, T \, T^T \,]$ and $k_{max}$ is the largest eigenvalue of $\vec{a}_{1 \, (2)} \vec{a}^{\,2}_{1 \, (2)} + T \, T^T$. For pure states, the geometric discord reduces to an entanglement measure.

\section{Ansatz for the density matrix}

A density operator that commutes with Dirac Hamiltonian can be used to derive entangling properties and open system effects of any Dirac-like system. A first step towards the derivation of such a density matrix can be implemented through the use of the explicit form of eigenspinors $u_s (p)$ and $v_s (p)$, the solutions to Dirac equation in momentum representation.
In an outstanding work, Wightman \cite{n028,n029} derived a formula for the density matrix of free spinor states with arbitrary polarization projections. That formula consists of a decomposition of the matrix $ u_s (p) u_s ^\dagger (p) \gamma_0$ in terms of the sixteen $\Gamma_i$ matrices and it has been used for very particular quantum field calculations. Despite resulting into normalized pure states, the Wightman procedure could not be straightforwardly generalized to obtaining spinors when the dynamics include external field potentials, which supports the main proposal of our work.

To quantify the role of external potentials presented in Table \ref{Tab:01} onto the \textit{spin-parity} entanglement of Dirac states, our preliminary analysis only accounts for constant external fields with $\vec{E} = 0$, as to have the Hamiltonian from (\ref{E04T}) rewritten as
\begin{equation}
\label{E04}
\tilde{H} = H - A^0\,I_4 = \gamma_0 m + \vec{\alpha} \cdot \vec{\mathcal{P}} + i \gamma_0 \gamma_5 \mu - \gamma_5 q + \gamma_5 \vec{\alpha} \cdot \vec{W} + i \chi_a \vec{\gamma} \cdot \vec{B} + \,\kappa_a\, \gamma_5 \vec{\gamma} \cdot \vec{B},
\end{equation}
where the mass (scalar) term is approached by replacing $m + \phi_S$ by $m$.
The inclusion of a non-vanishing electric field-like component into the above Hamiltonian can be performed by making the replacements $\chi_a \vec{B} \rightarrow \chi_a \vec{B} + \kappa_a \vec{E}$ and $\kappa_a \vec{B} \rightarrow \kappa_a \vec{B} - \chi_a \vec{E}$.

Instead of solving the Dirac equation with an arbitrary potential matrix, one may adopt another step-by-step approach over $\tilde{H}$, as to derive pure state density matrix for Dirac bi-spinors.
Once introduced the modified Hamiltonian, $\tilde{H}$, such that, $\Tr[\tilde{H}] = 0$ one has
\begin{eqnarray}
\label{C02}
\tilde{H}^2 &=& c_1\, I_4 + 2 \mathcal{O}, \nonumber \\
\frac{(\tilde{H}^2 - c_1\, I_4)^2}{4} &=& c_2\, I_4 + 2 [\, (\mu \chi_a - m \,\kappa_a\,)\, (\vec{W} \cdot \vec{B}) - q (\vec{\mathcal{P}} \cdot \vec{W})\,] \tilde{H},
\end{eqnarray}
where
\begin{eqnarray}
c_1 &=& \frac{1}{4}\Tr[\tilde{H}^{2}], \nonumber \\
c_2 &=& \frac{1}{16}\Tr\left[\left(\tilde{H}^2- \frac{1}{4}\Tr[\tilde{H}^{2}]\right)^{2}\right],
\end{eqnarray}
and $\mathcal{O}$ is a traceless operator given by
\begin{eqnarray}
\mathcal{O} &=& \vec{\Sigma} \cdot [ \,  (\mu \chi_a - m \,\kappa_a\,) \vec{B} - q \vec{\mathcal{P}} \, ] + \gamma_0 \vec{\Sigma} \cdot [\, m \vec{W} + \chi_a \vec{\omega}_B \,] + i \gamma_0 \gamma_5 \vec{\Sigma} \cdot [\, \mu \vec{W} + \,\kappa_a\, \vec{\omega}_B\,]  \nonumber \\
&-& q \gamma_5 \vec{\Sigma} \cdot \vec{W} + (\vec{\mathcal{P}} \cdot \vec{W})\gamma_5 - \,\kappa_a\, (\vec{W}\cdot \vec{B})\gamma_0 + i \chi_a (\vec{W} \cdot \vec{B})\gamma_0 \gamma_5.
\end{eqnarray}
with
\begin{eqnarray}
\label{C05}
\vec{\omega}_B &=& \vec{\mathcal{P}}\times \vec{B}\nonumber\\
\vec{\Sigma} &=& \mbox{diag}\{\vec{\sigma},\vec{\sigma}\}\nonumber\\
c_1 &=& \mathcal{P}^2 + m^2 + \mu^2 + q^2 + W^2 + (\,\kappa_a^2 + \chi_a^2) B^2, \nonumber \\
c_2 &=& [ (\mu \chi_a - m \,\kappa_a\,)\vec{B} - q \vec{\mathcal{P}}]^2 + [m \vec{W} + \chi_a \vec{\omega}_B]^2 + [\mu \vec{W} + \,\kappa_a\, \vec{\omega}_B]^2 + \nonumber \\
 &+& q^2 W^2 + (\vec{\mathcal{P}} \cdot \vec{W})^2 + (\,\kappa_a^2 + \chi_a)^2 (\vec{W} \cdot \vec{B})^2.
\end{eqnarray}

For such a generalized form of the Hamiltonian, it is easy to notice that the \textit{ansatz}
\begin{equation}
\label{C01}
\varrho = \frac{1}{4} \left( \, I_4 + \frac{(-1)^s}{\sqrt{c_2}} \mathcal{O} \, \right) \left(\, I_4 + \frac{(-1)^n}{\vert \, \lambda \, \vert} \tilde{H} \, \right),
\end{equation}
with $n,s = \{1,2\}$, provides the stationary solutions of the Liouville equation. The parameter $\lambda$ is related to the energy eigenvalues and it is calculated through the equation
\begin{equation}
\frac{(\lambda^2 - c_1)^2}{4} = c_2 + 2 [\, (\mu \chi_a - m \,\kappa_a\,)\, (\vec{W} \cdot \vec{B}) - q (\vec{\mathcal{P}} \cdot \vec{W})\,] \lambda.
\end{equation}
If $\mathcal{O}^2 = c_2\, I_4$ then $\lambda$ is evaluated as
\begin{equation}
\label{C03}
\lambda = (-1)^n \sqrt{ c_1 + 2 \,(-1)^s \sqrt{c_2}},
\end{equation}
and it is the re-defined mean energy of ($\ref{C01}$),
\begin{equation}
\label{C04}
E=Tr[\tilde{H} \varrho] =\lambda.
\end{equation}
The condition $\mathcal{O}^2 = c_2\, I_4$ is accomplished by an adequate choice of parameters.
For instance, by the identification of some relative orientation of the vectors $\vec{\mathcal{P}}$, $\vec{W}$ and $\vec{B}$ for which $\varrho$ is a pure state.
On the other hand, if $\mathcal{O} = 0$ then one has
\begin{equation}
\mbox{Tr}[\varrho^2] = \frac{1}{4} \left(\, 1 + \frac{c_1}{\lambda^2} \, \right),
\end{equation}
and (\ref{C01}) is a mixed state.

For pure states, entanglement properties are given in terms of the Bloch vectors, as described in Sec. II. If $\mathcal{O}^2 = c_2 \, I_4$, the Bloch vector of the parity subsystem of (\ref{C01}) reads
\begin{eqnarray}
\label{C06}
\vec{a}_2 &=& \frac{(-1)^s}{\sqrt{c_2}} [ (\mu \chi_a - m \,\kappa_a\,) \vec{B} - q \vec{\mathcal{P}}] + \frac{(-1)^n}{\vert \lambda \vert} \vec{W} + \frac{(-1)^{s+n}}{\vert \lambda \vert \sqrt{c_2}} [ \, m (m \vec{W} + \chi_a \vec{\omega}_B) \nonumber \\
&+& \mu(\mu \vec{W} + \,\kappa_a\, \vec{\omega}_B) + (\vec{\mathcal{P}} \cdot \vec{W}) \vec{\mathcal{P}} + q^2 \vec{W} + (\,\kappa_a^2 + \chi_a^2)(\vec{W} \cdot \vec{B})\vec{B} \, ].
\end{eqnarray}
From now on one may call it simply the Bloch vector and drop the subscript $2$. The above formula is as general as possible and applications require a fine specification of parameters, i. e. the numerical values of $m$, $\mu$ and $q$,  the modulus of the vectors $\vec{\mathcal{P}}$, $\vec{B}$, $\vec{W}$ and their relative orientations.
In the following, one shall identify four specific frameworks, and for all of them, time and space components of the vector potential ($A^0$ and $\vec{A}$) are implicitly included into the respective definitions of $\tilde{H}$ and of $\vec{\mathcal{P}}$ as that they transform as those ones under Poincar\'e symmetries.

\subsection*{Case $\vec{W} = \vec{B} = 0$ and $q = 0$.}

\begin{figure}[h]
\includegraphics[width = 7.5 cm]{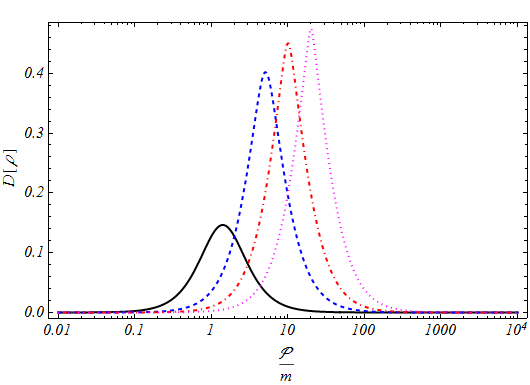}
\caption{Geometric discord for the mixed state obtained through the \textit{ansatz} for the Hamiltonian with pseudoscalar potential as function of $\mathcal{P}/m$ for $\mu/m = 1$ (continuous line), $5$ (dashed line), $10$ (dot-dashed line) and $20$ (dotted line). Geometric discord reaches its maximum value for $\mathcal{P}^2  = m^2 + \mu^2 $. Even for large values of $\mathcal{P}/m$ quantum correlations are exhibited if $\mu/m$ is sufficient large.}
\label{fig:01}
\end{figure}

As a first approach, let one consider the Dirac Hamiltonian of a free particle, modified by the inclusion of a pseudoscalar interaction,
\begin{equation}
\label{E04a}
\tilde{H} = \gamma_0 m + \vec{\alpha} \cdot \vec{\mathcal{P}} + i \gamma_0 \gamma_5 \mu,
\end{equation}
where hereafter the role of $A^0$ can be completely neglected.
In this exceptional case, the operator $\mathcal{O}$ is null and therefore the state obtained through the {\em ansatz} from (\ref{C01}), $\varrho_{PS}$, is mixed.
Given that its concurrence is null for any value of $\mu$, $m$ and $\vec{\mathcal{P}}$, this state is separable. Quantum correlations other than the entanglement are exhibited and quantified by the geometric discord computed from (\ref{disc}) and evaluated as
\begin{equation}
\label{A02}
\mathcal{D}[\varrho_{PS}] = \frac{1}{2} - \sqrt{\frac{1}{4} - \frac{\mu^2 \mathcal{P}^2}{\lambda_{PS} ^4}},
\end{equation}
where
\begin{equation}
\label{C03B}
\lambda_{PS} = (-1)^n \sqrt{\mathcal{P}^2 + m^2 + \mu^2}.
\end{equation}
Fig.~\ref{fig:01} shows $\mathcal{D}[\varrho_{PS}]$ as function of $\mathcal{P}/m$. The geometric discord is null if $\mu = 0$, i.e. for free particles, and it reaches a maximized value at $\mathcal{P} = \sqrt{m^2 + \mu^2}$ vanishing if $ \mathcal{P}/\mu \ll 1$ or $ \mathcal{P}/\mu \gg 1$. Quantum correlations do not vanish in the ultra-relativistic (UR) limit if the pseudoscalar potential is sufficiently large.

In the absence of the pseudoscalar potential the Hamiltonian exhibits a bi-spinor free particle(-like) structure, for which $\tilde{H} = \tilde{H}_0 = \vec{\mathcal{P}} \cdot \vec{\alpha} + m \beta$ and the density operator obtained through (\ref{C01}) is
\begin{equation}
\varrho_0^{\{n\}}(\vec{\mathcal{P}},m) = \frac{1}{4} \left[1 + \frac{(-1)^n}{\sqrt{\mathcal{P}^2 + m^2}} (\vec{\mathcal{P}} \cdot \vec{\alpha} + m \beta) \right],
\end{equation}
This density matrix can be related to the polarized spinors $u_s (\mathcal{P})$ and $v_s(\mathcal{P})$ (see \cite{n028}) through the projection of $\varrho_0$ into polarization states,
\begin{eqnarray}
v_s (\vec{\mathcal{P}}) v_s ^\dagger (\vec{\mathcal{P}}) &=& - 2 \sqrt{\mathcal{P}^2 + m^2}(1 - s \gamma_5 z_\mu \gamma^\mu) \, \varrho_0 ^{\{0\}} (- \vec{\mathcal{P}}, m), \nonumber \\
u_s (\vec{\mathcal{P}}) u_s ^\dagger (\vec{\mathcal{P}}) &=& - 2 \sqrt{\mathcal{P}^2 + m^2} (1 - s \gamma_5 z_\mu \gamma^\mu) \, \varrho_0 ^{\{1\}} (- \vec{\mathcal{P}}, m).
\end{eqnarray}
The state $\varrho_0 ^{\{n\}}$ corresponds to an unpolarized state that can be converted into a polarized state along the direction $z_\mu$, by the inclusion of the operator $(1 - s \gamma_5 z_\mu \gamma^\mu)$.

\subsection*{Case $\chi_a = q= 0$ and $\vec{W} = 0$.}

Let one now consider the free particle(-like) Hamiltonian ($\tilde{H}$ with $\mathcal{P}$) from (\ref{E04}) only with the inclusion of tensor and pseudoscalar potentials, i. e. with $\chi_a = q = 0$ and $\vec{W} = 0$.
\begin{equation}
\label{E04b}
\tilde{H} = \gamma_0 m + \vec{\alpha} \cdot \vec{\mathcal{P}}  + i \gamma_0 \gamma_5 \mu + \,\kappa_a\, \gamma_5 \vec{\gamma} \cdot \vec{B},
\end{equation}

Such system can include effects of the magnetic momentum non-minimal coupling resulted from second order correction diagrams of quantum electrodynamics.
Under the above assumptions one derives the following expressions for $c_1$ and $c_2$,
\begin{eqnarray}
\label{G01}
c_1 &=& \mathcal{P}^2 + m^2 + \mu^2 + \,\kappa_a^2 B^2 \nonumber \\
c_2 &=&\,\kappa_a^2 (m^2 B^2 + \omega_B ^2),
\end{eqnarray}
and for the Bloch vector,
\begin{equation}
\label{G02}
\vec{a} = \frac{(-1)^s\,\kappa_a\,}{\sqrt{c_2}} \left[ m \vec{B} - \frac{(-1)^n}{\vert \lambda_a \vert}\mu \,\vec{\omega}_B \right].
\end{equation}
with
\begin{equation}
\label{C03B}
\lambda_{a} = (-1)^n \sqrt{\mathcal{P}^2 + m^2 + \mu^2 + \kappa_a ^2 B^2 + 2 (-1)^s \sqrt{m^2 \kappa_a^2 B^2 + \kappa_a^2 \omega_B^2}}.
\end{equation}

\begin{figure}[h]
\includegraphics[width = 7.5 cm]{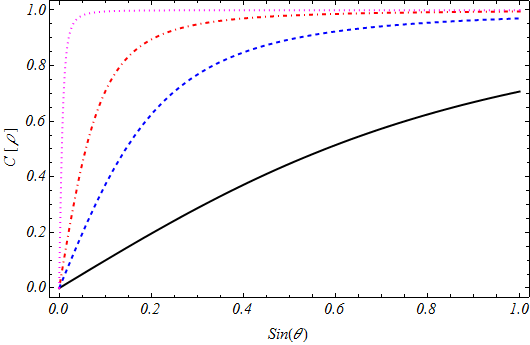}
\caption{Concurrence for the state associated to pseudoscalar and tensor potentials as function of $\sin \theta$ for $\mu/m = 0$, and $\mathcal{P} /m = 1.0$ (continuous line), $4.0$ (dashed line), $10.0$ (dot-dashed line) and $100.0$ (dotted line). Entanglement vanishes in the non-relativistic limit ($\mathcal{P}/m \rightarrow 0$) and for increasing values of $\mathcal{P}/m$ the entanglement tends to a square-shaped (Heaviside theta) function.}
\label{fig:02}
\end{figure}

By specifying some values of the above involved parameters, if the pseudo-scalar potential is absent ($\mu = 0$), the entanglement depends only on $\mathcal{P} /m$ and on the relative orientation between $\vec{B}$ and $\vec{\mathcal{P}}$ parameterized by their relative angle $\theta$. Fig.~\ref{fig:02} depicts the concurrence in terms of $\sin \theta$ for $\mu = 0$ and for pertinent values of $\mathcal{P}/m$. Concurrence is an increasing function of $\theta$ and, once the UR limit is reached, the entanglement tends to the Heaviside theta function $\Theta(x)$. Therefore, if $\mathcal{P}/m$ goes to $\infty$ and $\vec{\mathcal{P}}$ is not parallel to the magnetic field $\vec{B}$, then the state is maximally entangled. Conversely, in the non-relativistic limit, $\mathcal{P}/m$ goes to $ 0$, the concurrence vanishes.

\begin{figure}[h]
\includegraphics[width = 8.1 cm]{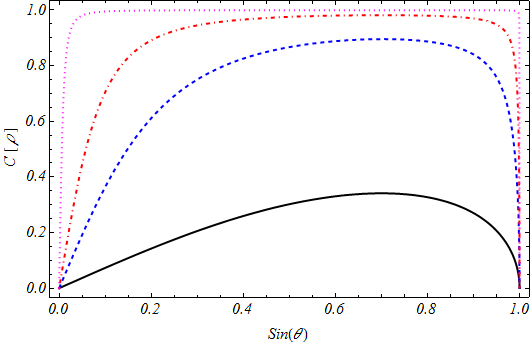}
\includegraphics[width = 8 cm]{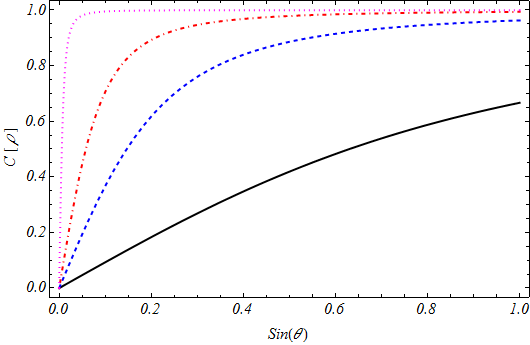}
\caption{Concurrence for the state associated to pseudoscalar and tensor potentials ($\sim$ Fig.~\ref{fig:02}) with $B = \sqrt{\mathcal{P}^2 + m^2}$, as function of $\sin \theta$. The values of $\mathcal{P}/m$ and the plot lines follow the correspondence with those ones from Fig.~\ref{fig:02}. The entanglement for $s = 1$ vanishes at $\theta = \pi/2$, and for $s=2$ the results behave like those shown in Fig.~\ref{fig:02}. In the UR limit the entanglement tends to a step function.}
\label{fig:03}
\end{figure}

For $\theta = 0$, the concurrence is null even for $\mu\neq 0$. Otherwise, if $\vec{\mathcal{P}}$ and $\vec{B}$ are orthogonal, the entanglement also vanishes $\mu\neq 0$ only if $s = 1$ and
\begin{equation}
\label{G03}
\,\kappa_a\, B = \sqrt{\mathcal{P}^2 + m^2}.
\end{equation}
Such a particular configuration has the concurrence depicted in Fig.~\ref{fig:03} for both $s = 1$ and $s = 2$. For the first case the entanglement has a maximal point and it vanishes at $\theta = 0$ and $\theta = \pi/2$. In the UR limit, the concurrence tends to a square function $\Theta(x) - \Theta(x - 1)$. For $x \neq 0,1$ the state is maximally entangled. Finally, for $s=2$, the entanglement becomes an increasing function of $\sin \theta$ and the results are similar to those for $\mu = 0$.

An analogous configuration of external field is implemented by setting $q = \,\kappa_a\, = 0$ and $\vec{W}=0$, i. e. under the inclusion of only pseudotensor and pseudoscalar potentials into the Dirac Hamiltonian from (\ref{E04}), as to have
\begin{equation}
\label{E04c}
\tilde{H} = \gamma_0 m + \vec{\alpha} \cdot \vec{\mathcal{P}} + i \gamma_0 \gamma_5 \mu + i \chi_a \vec{\gamma} \cdot \vec{B}.\end{equation}
In this framework, the expressions for $c_{1,2}$ and $\vec{a}$ are the same as those from Eqs.~(\ref{G01}) and (\ref{G02}) by replacing $\mu$ by $m$ and $\kappa_a$ by $\chi_a$.

\subsection*{Case $\vec{B} = 0$.}

Only pseudovector and pseudoscalar potentials are added to the free particle(-like) Hamiltonian ($\tilde{H}$ with $\mathcal{P}$) when $\vec{B} = 0$. In this case, one has
\begin{equation}
\label{E04d}
\tilde{H} = \gamma_0 m + \vec{\alpha} \cdot \vec{\mathcal{P}} + i \gamma_0 \gamma_5 \mu - \gamma_5 q + \gamma_5 \vec{\alpha} \cdot \vec{W},
\end{equation}
and the relevant parameters are given by
\begin{eqnarray}
c_1 &=& \mathcal{P}^2 + M^2 + q^2+ W^2, \nonumber \\
c_2 &=& q^2 \mathcal{P}^2 + (M^2 + q^2) W^2 + (\vec{\mathcal{P}}\cdot \vec{W})^2, \nonumber \\
\vec{a} &=& \frac{(-1)^s \, q \, \vec{\mathcal{P}}}{\sqrt{c_2}} + \frac{(-1)^n \, \vec{W}}{\vert \lambda_{aB} \vert} + \frac{(-1)^{s+n}}{\vert \lambda_{aB} \vert \sqrt{c_2}} \left[(M^2 + q^2) \vec{W} + (\vec{\mathcal{P}}\cdot\vec{W})\vec{\mathcal{P}} \right],
\end{eqnarray}
where $M^2 = m^2 + \mu^2$ and
with
\begin{equation}
\label{C03B}
\lambda_{aB} = (-1)^n \sqrt{\mathcal{P}^2 + M^2 + q^2 + W^2 + 2 (-1)^s \sqrt{q^2 \mathcal{P}^2 + (M^2 + q^2) W^2 + (\vec{\mathcal{P}}\cdot \vec{W})^2}}.
\end{equation}
which is obtained by setting either $q = 0$ or $\vec{W}\cdot\vec{\mathcal{P}} = 0$ as to be consistent with Eq.~(\ref{C03}).

\begin{figure}[h]
\includegraphics[width = 8.1 cm]{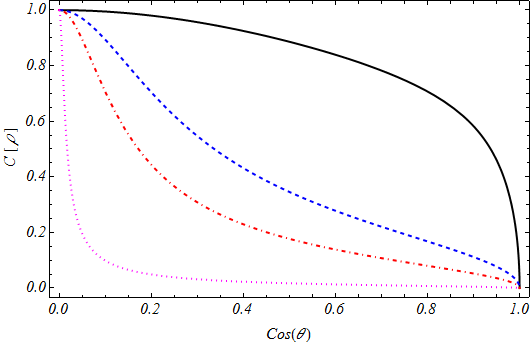}
\includegraphics[width = 8 cm]{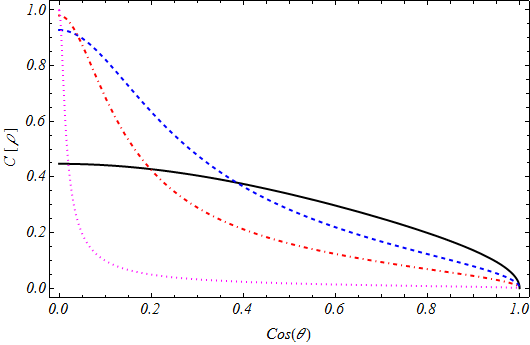}
\includegraphics[width = 8.1 cm]{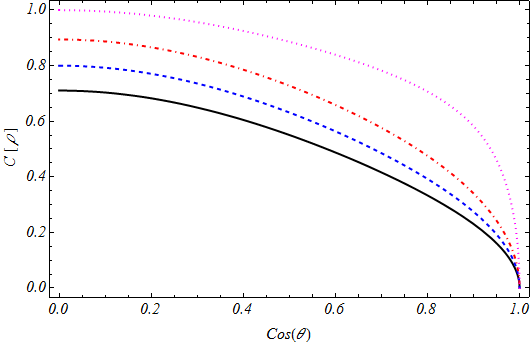}
\includegraphics[width = 8 cm]{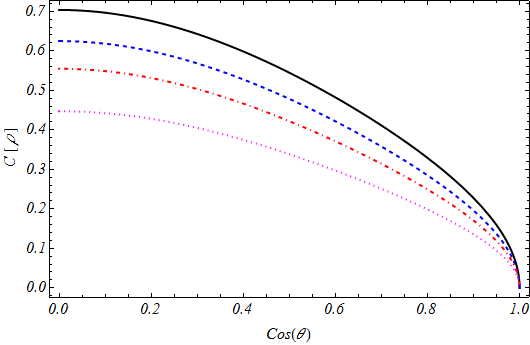}
\caption{Concurrence for the state associated to pseudoscalar and pseudovector potentials as function of $ \cos \theta$, where $\theta$ is the angle between the spatial component of the pseudovector potential, $\vec{W}$, and the canonical momentum $\vec{\mathcal{P}}$. For the first row, one has $W/ M = 1$ and $\mathcal{P}/M = 1$ (continuous line), $5$ (dashed line), $10$ (dot dashed line) and $100$ (dotted line). For the second row, one has $\mathcal{P}/M = 1$ and $W/M = 0.01$ (continuous line), $0.25$ (dashed line), $0.5$ (dot dashed line) and $1$ (dotted line). The left column corresponds to $s = 1$ and the right column to $s=2$. One notices that the concurrence is a decreasing function of $\cos \theta$, which vanishes for $\vec{\mathcal{P}}$ perpendicular to $\vec{W}$. For $\mathcal{P}/M \rightarrow \infty$ the degree of entanglement is suppressed for $\theta \neq 0$. }
\label{fig:04}
\end{figure}

\begin{figure}[h]
\includegraphics[width = 8.1 cm]{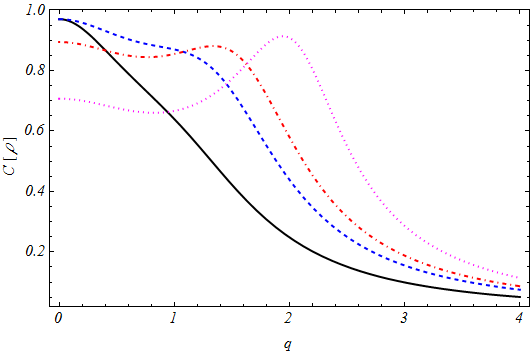}
\includegraphics[width = 8 cm]{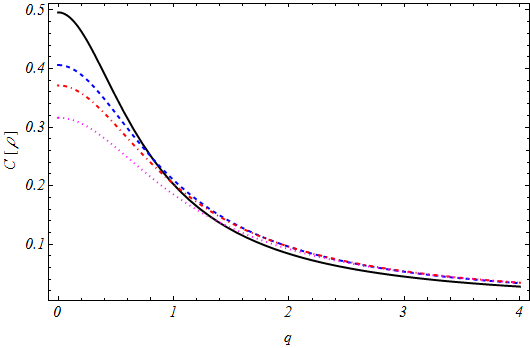}
\caption{Concurrence for the state associated to pseudoscalar and pseudovector potentials as function of the time component of the pseudovector potential, $q$, for $\mathcal{P}/ M = 1$ and $W/M = 0.75$ (continuous line), $1.25$ (dashed line), $1.5$ (dot dashed line) and $2$ (dotted line). The left plot corresponds to $s = 1$ and the right plot to $s = 2$. For $s=2$, the entanglement strictly decreases with $q$. Otherwise, for $s=1$, the entanglement exhibits maximum values that are more evident for high values of $W/M$. In both cases concurrence vanishes for $q/M \rightarrow \infty$.}
\label{fig:05A}
\end{figure}

For $q = 0$, the entanglement is a decreasing function of $\cos \theta = \vec{W}\cdot\vec{\mathcal{P}}/W \mathcal{P}$, as depicted in the first row of Fig.~\ref{fig:04} for several values of $\mathcal{P}/M$, with $W/M$ fixed (first row) and for several values of $W/M$, with $\mathcal{P}/M$ fixed (second row). For the field $\vec{W}$ parallel to $\vec{\mathcal{P}}$ the state is continuously separable. On the other hand, the entanglement is maximized for $\theta = \pi/2$ only if $W/m = 1$. For a fixed $\mathcal{P}/M$ the entanglement decreases with $W/M$ for $s = 1$ and it increases for $s =2$ (c. f. Fig.~\ref{fig:04}).
Conversely, for $\vec{W}\cdot\vec{\mathcal{P}}=0$ and non-vanishing values of $q$, the concurrence exhibits maximum values as shown by Fig.~\ref{fig:05A}. As $W/M$ increases for a fixed value of $\mathcal{P}/M$, the entanglement reaches a maximum point for $s =1$ and it vanishes for large values of $q/M$. For $s = 2$, the concurrence is a decreasing function of $q$.

\subsection*{Case $\chi_a = 0$.}

\begin{figure}[h]
\includegraphics[width = 8.1 cm]{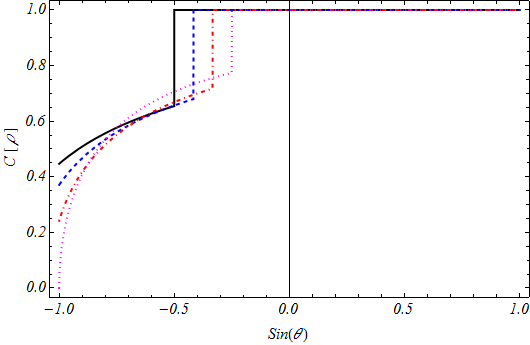}
\includegraphics[width = 8 cm]{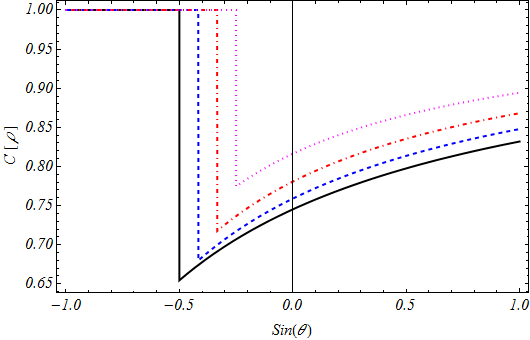}
\includegraphics[width = 8.1 cm]{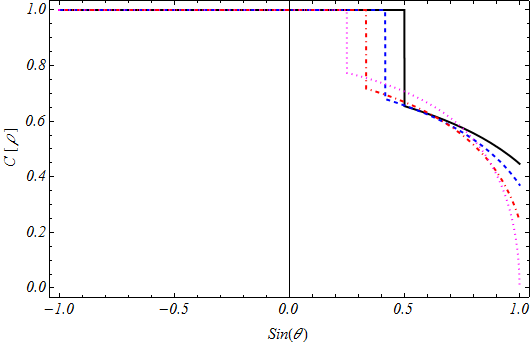}
\includegraphics[width = 8 cm]{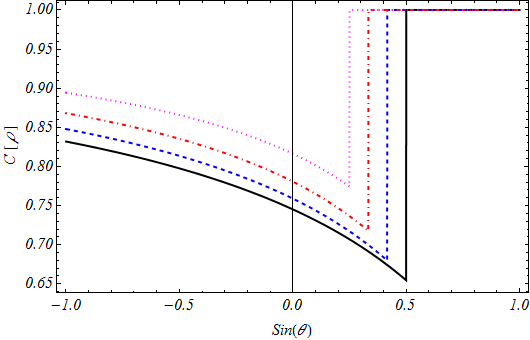}
\caption{Concurrence for the state associated to pseudoscalar, tensor and pseudovector external fields as function of $\sin{\theta}$, where $\theta$ is the angle between $\vec{B}$ and $\vec{\mathcal{P}}$, for $\vec{W}$ parallel to $\vec{\omega}_B$ (first row) and for $\vec{W}$ antiparallel to $\vec{\omega}_B$ (second row). The left column corresponds to $s = 1$, the right column to $s = 2$ and the plots are for $\mathcal{P}/\mu = 1$ (continuous line), $1.2$ (dashed line), $1.5$ (dot dashed line) and $2$ (dotted line). Assuming that $W / \mu = 1$,  one notices a discontinuity in concurrence for $\vert \sin \theta \vert = \vert \sin \theta_c \vert = \mu W / \mathcal{P} B$. This discontinuity corresponds to a transition from partially to maximally entangled states.}
\label{fig:06}
\end{figure}

By including pseudovector and tensor potentials into the free particle(-like) Hamiltonian ($\tilde{H}$ with $\mathcal{P}$) one has
\begin{equation}
\label{E04e}
\tilde{H} = \gamma_0 m + \vec{\alpha} \cdot \vec{\mathcal{P}} + i \gamma_0 \gamma_5 \mu - \gamma_5 q + \gamma_5 \vec{\alpha} \cdot \vec{W} +  \gamma_5 \vec{\gamma} \cdot \vec{B},
\end{equation}
where it has been set $\kappa_a = 1$, in order to not overcharge the notation, and the relevant parameters are given by
\begin{eqnarray}
c_1 &=& \mathcal{P}^2 + m^2 + \mu^2 + q^2 + W^2 + \, B^2, \nonumber \\
c_2&=& (m \, \vec{B} + q \vec{\mathcal{P}})^2 + (m^2 + \mu^2 + q^2)W^2 \nonumber \\
&+& 2 \mu \,\vec{W}\cdot \vec{\omega}_B + \,\omega_B^2 + (\vec{\mathcal{P}} \cdot \vec{W})^2 + \,(\vec{W} \cdot \vec{B})^2, \nonumber \\
\vec{a} &=& \frac{(-1)^{s+1}}{\sqrt{c_2}}\,(m \,\vec{B} + q \vec{\mathcal{P}}) + \frac{(-1)^n}{\vert \lambda_{a\delta} \vert} \vec{W} \nonumber \\
&+& \frac{(-1)^{s+n}}{\vert \lambda_{a \delta} \vert \sqrt{c_2}}\, \left[ (m^2 + \mu^2 + q^2)\vec{W} + \,\vec{\omega}_B + (\vec{\mathcal{P}} \cdot \vec{W}) \vec{\mathcal{P}} + \,(\vec{W} \cdot \vec{B})\vec{B} \right].
\end{eqnarray}
with
\begin{eqnarray}
\label{C03D}
\lambda_{a\delta} &=& (-1)^n \, \biggl[\, \mathcal{P}^2 + m^2 + \mu^2 + q^2 + W^2 + B^2   \\
 &+& 2(-1)^s \sqrt{m^2 B^2 + (m^2 + \mu^2 + q^2)W^2 + 2 \mu (\vec{W}\cdot \vec{\omega}_B) +(\vec{\mathcal{P}}\cdot \vec{W})^2+ (\vec{W}\cdot \vec{B})^2 }\, \, \biggr]^{1/2}, \nonumber
\end{eqnarray}
which is obtained under the constraint
\begin{equation}
m \,\vec{W}\cdot\vec{B} + q \vec{\mathcal{P}}\cdot \vec{W} = 0,
\end{equation}
as to fit Eq.~(\ref{C03}).
One way to accomplish the above condition is by choosing $q = 0$ and $m = 0$. This simple framework describes a massless fermion under the influence of a tensor field and a pseudovector external field that has no time-like component. In this case the entangling properties are obtained in terms of
\begin{eqnarray}
c_1 &=& \mathcal{P}^2  + \mu^2 + W^2 + \,B^2, \nonumber \\
c_2 &=& (\mu \vec{W} + \, \vec{\omega}_B)^2  + (\vec{\mathcal{P}}\cdot \vec{W})^2 + \,(\vec{W}\cdot\vec{B})^2,
\end{eqnarray}
and the Bloch vector
\begin{equation}
\vec{a} = \frac{(-1)^n}{\vert \lambda_{a\delta} \vert} \left[ \vec{W} + \frac{(-1)^s}{\sqrt{c_2}} \left( \mu (\mu \vec{W} + \vec{\omega}_B) + (\vec{\mathcal{P}}\cdot\vec{W})\vec{\mathcal{P}} + \,(\vec{W} \cdot \vec{B}) \vec{B} \right) \right].
\end{equation}
For $\vec{B}$ and $\vec{\mathcal{P}}$ perpendicular to $\vec{W}$, the expressions for $c_1$ and $c_2$ simplify to
\begin{eqnarray}
\label{F03}
c_1 &=& \mu^2 + \mathcal{P}^2 + W^2 + \, B^2, \nonumber \\
c_2 &=& (\mu W \pm \,\mathcal{P} B \sin\theta)^2,
\end{eqnarray}
where the the minus (plus) sign stands for $\vec{W}$ (anti)parallel to $\vec{\omega}_B$. The modulus of the Bloch vector is
\begin{equation}
a^2 = \frac{1}{\lambda_{a\delta}^2}\left[\mu^2 + W^2 + 2 (-1)^s \mu W \mbox{sign}({\mu W \pm \,\mathcal{P} B \sin\theta})\right],
\end{equation}
with
\begin{equation}
\label{C03D}
\lambda_{a\delta} = (-1)^n \bigl[ \,\,  \mathcal{P}^2 + \mu^2 + W^2 + \, B^2 + 2 (-1)^s \vert \, \mu W \pm \,\mathcal{P} B \sin\theta \, \vert \,\,  \bigl]^{1/2}.
\end{equation}
If $\mu \,W < \mathcal{P} B$ this expression exhibits an abrupt change of behavior at $\sin\theta_c = \mu \,W / \mathcal{P} B$, in terms of this quantity the above equation reads
\begin{eqnarray}
\label{F04}
a^2 &=& \frac{1}{\lambda_{a\delta}^2} (\, W + (-1)^s \mbox{sign}[\sin \theta_c \pm \sin \theta] \mu \,)^2.
\end{eqnarray}
with
\begin{equation}
\label{C03D}
\lambda_{a\delta} = (-1)^n \bigl[ \,\,  \mathcal{P}^2 + \mu^2 + W^2 + \, B^2 + 2 (-1)^s \mathcal{P} B  \vert \, \sin \theta_c \pm \, \sin\theta \, \vert \,\,  \bigl]^{1/2}.
\end{equation}
The behavior changes at $\sin \theta = \mp \sin \theta_c$, with the minus (plus) sign for $\vec{W}$ (anti)parallel to $\vec{\omega}_B$. From (\ref{F03}), one concludes that $\vec{W}$ antiparallel to $\vec{\omega}_B$ corresponds to the converse case with the replacement of $\theta$ by $-\theta$. For $\vec{W}$ parallel to $\vec{\omega}_B$, the expression (\ref{F04}) simplifies to
\begin{eqnarray}
\label{F01}
a^2&=& \frac{1}{\lambda_{a\delta}^2} \left[W + (-1)^{s+1} \mu \right]^2 \,\,\,\,\, \mbox{for  } x < -x_c, \nonumber \\
&=& \frac{1}{\lambda_{a\delta}^2} \left[W + (-1)^{s} \mu \right]^2 \,\,\,\,\, \mbox{for  } x > -x_c,
\end{eqnarray}
therefore, if $W/\mu = 1$ the state is maximally entangled for $s = 1$ if $x>-x_c$, and for $s = 2$ if $x<-x_c$.
The corresponding concurrence for $W/\mu = 1$ and for four values of $\mathcal{P}/\mu$ are depicted in Fig.~\ref{fig:06}.
The entanglement is an increasing function of $\sin \theta$, however, it exhibits a discontinuity at $\sin \theta = - \sin \theta_c$ where there is an abrupt transition to a maximally entangled state.
The same behavior is expected for $\vec{W}$ antiparallel to $\vec{\omega}_B$ but with $\theta$ replaced by $- \theta$. In this case, the entanglement is a decreasing function of $\sin\theta$,  as shown in the second row of Fig.~\ref{fig:06}.

\begin{figure}[h]
\includegraphics[width = 8.1 cm]{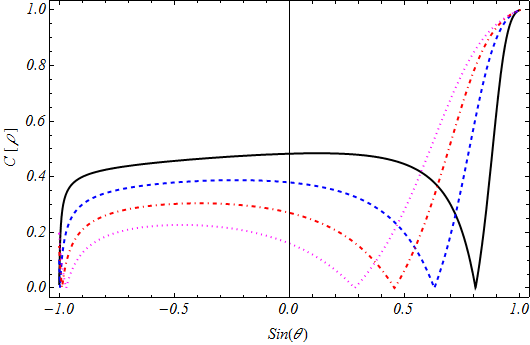}
\includegraphics[width = 8 cm]{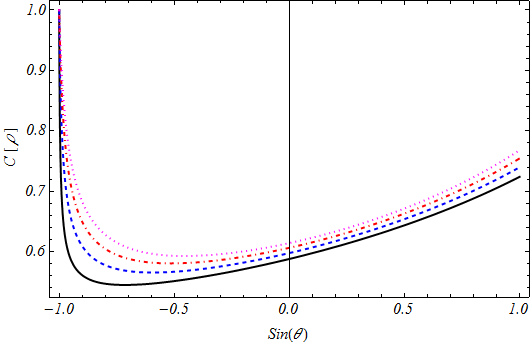}
\includegraphics[width = 8.1 cm]{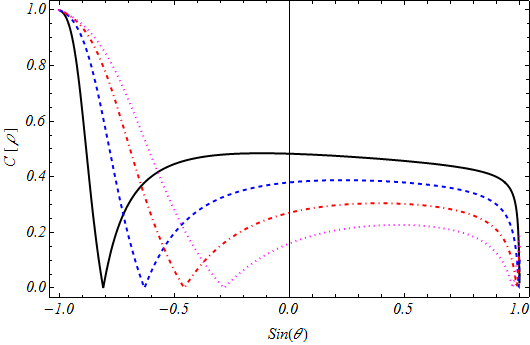}
\includegraphics[width = 8 cm]{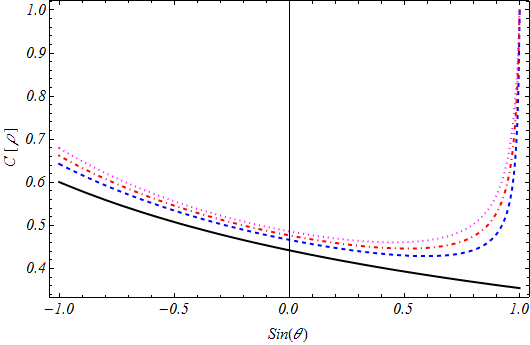}
\caption{Concurrence for the state associated to pseudoscalar, tensor and pseudovector external fields as function of $\sin{\theta}$, where $\theta$ is the angle between $\vec{W}$ and $\vec{\mathcal{P}}$, for $\vec{B}$ parallel to $\vec{\omega}_W$ (first row) and for $\vec{B}$ antiparallel to $\vec{\omega}_W$ (second row). The plots are for $s = 1$ (left column) and $s = 2$ (right column). The values of $\mathcal{P}/\mu$ and the plot-styles are in correspondence with Fig.~\ref{fig:06}.
Assuming that $W / \mu = 1$, one notices that, for $s = 1$, the state has two zero points, which depend on the rapports $\mathcal{P}/\mu$ and $W/\mu$.}
\label{fig:08}
\end{figure}

In case of considering $\vec{B}\cdot\vec{\mathcal{P}} = \vec{W}\cdot\vec{B}=0$, i. e. for $\vec{B}$ orthogonal to $\vec{\mathcal{P}}$ and $\vec{W}$, the expression for $c_1$ keeps as it stands and $c_2$ and the Bloch vector are rewritten as
\begin{eqnarray}
c_2 &=& (\mu \vec{W} + \vec{\omega}_B)^2 + (\vec{\mathcal{P}} \cdot \vec{W})^2, \nonumber \\
\vec{a} &=& \frac{(-1)^n}{\vert \lambda_{a\delta} \vert}\left[ \vec{W} + \frac{(-1)^s}{\sqrt{c_2}}( \mu(\mu \vec{W} + \vec{\omega}_B) + (\vec{\mathcal{P}} \cdot \vec{W}) \vec{\mathcal{P}} )\right].
\end{eqnarray}
The entanglement depends on the orientation of $\vec{B}$ relative to the plane where $\vec{\mathcal{P}}$ and $\vec{W}$ lie. Fig.~\ref{fig:08} shows the concurrence as function of $\sin \theta$, for $\theta$ corresponding to the angle between $\vec{W}$ and $\vec{\mathcal{P}}$, for $\vec{B}$ parallel and antiparallel to $\vec{\omega}_W$. For the values adopted therein, the entanglement has two zero points for $s=1$ and one minimum point for $s=2$.

\begin{figure}[h]
\includegraphics[width = 8.1 cm]{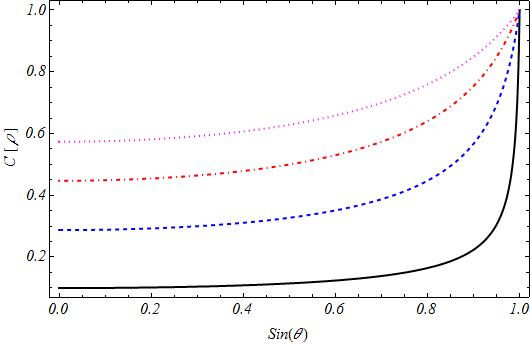}
\caption{UR limit for the concurrence of the state associated to pseudoscalar, tensor and pseudovector potentials as function of $\sin{\theta}$ for $B/W = 0.1$  (continuous line), $0.3$ (dashed line), $0.5$ (dot dashed line) and $0.7$ (dotted line). If $\vec{\mathcal{P}}$ and $\vec{W}$ are perpendicular, the state is maximally entangled.}
\label{fig:10}
\end{figure}

In the UR limit, the entanglement neither vanishes nor tends to its maximum value. For this case
\begin{equation}
a^2(\mathcal{P}/\mu \rightarrow \infty) = \frac{W^2 (1 - \sin^2 \theta)}{B^2 + W^2(1 - \sin^2 \theta)},
\end{equation}
and the state is maximally entangled only for $\theta = \pi/2$. Fig.~\ref{fig:10} shows the concurrence as function of $\sin \theta$ in the UR limit for several values of $B/W$. Entanglement is an increasing function of $B/W$, and for $B \gg W$ the state is maximally entangled for any value of $\sin \theta$.

Finally, if one considers $\kappa_a= 0$ and $\mu = 0$ analogous results are obtained by replacing $\mu$ by $m$, where the same entangling properties and effects are observed.

\section{Discussion and Conclusions}

Recent assertive results for which Dirac bi-spinors are described in terms of $\mbox{SU}(2) \otimes \mbox{SU}(2)$ {\em parity-spin} entangled states have been extended to $\mbox{SU}(2) \otimes \mbox{SU}(2)$ bi-spinor states driven by Poincar\'e classes of Dirac interacting potentials.
The generalized procedure developed in this work has allowed one to identify a suitable \textit{ansatz} for the density operator describing solutions to the Dirac Hamiltonian with external fields.
Consequently, Dirac-like global potentials driven by (pseudo)scalar, (pseudo)vector and tensor interactions have been identified as the drivers of $\mbox{SU}(2) \otimes \mbox{SU}(2)$ entanglement/separability, as well as the creators of quantum correlations (as for instance, the entanglement of formation) of controllable systems with enormous physical appeal when mapped onto the bi-spinor structure.

The focus of our analysis was on computing entangling properties similar to those of \textit{spin-parity} entanglement, induced by external potentials.
The simplifying hypothesis of considering constant external fields and neglecting localization effects associated to a carefully engendered {\em ansatz} has led to the construction of density operators (c. f. Eq.~(\ref{C01})) of pure eigenstates of the Dirac Hamiltonian, $H$.
Given that the square of the operator $\mathcal O$ (see Eq.~(\ref{C02})) contains only elements proportional to the identity operator, the energy eigenvalues were obtained in terms of algebraical functions of well-defined parameters related to the coupling constants of interacting external fields.
For such Hamiltonians, $H$, entangling properties of these states are obtained via the concurrence, which is described in terms of $Tr[\tilde{H}^{2}]$, $Tr[(\tilde{H}^2- 1/4\,Tr[\tilde{H}^{2}])^{2}]$ (c. f. Eq.~(\ref{C02})) and the Bloch vector components. General expressions for $Tr[\tilde{H}^{2}]$, $Tr[(\tilde{H}^2- 1/4 \, Tr[\tilde{H}^{2}])^{2}]$ and the additional parameters and the corresponding correlation properties were computed and discussed in four different Poincar\'e classes of external potentials.

For a Hamiltonian containing only the pseudoscalar potential, the states given by the \textit{ansatz} (\ref{C01}) are separable mixed states. The evaluation of the geometric discord shows that such states have quantum correlations that are concentrated around $\mathcal{P} = \sqrt{m^2 + \mu^2}$. For free particles ($\mu=0$) our results are related to the free particle bi-spinors, $u_s (p)$ and $v_s (p)$, via projections onto polarized states.

When only tensor and pseudoscalar potentials are present, for the value of a magnetic field given by (\ref{G03}), entanglement vanishes whether $\vec{B}$ is either parallel or perpendicular to $\vec{\mathcal{P}}$. In addition, the entanglement tends to a square-shaped function in the UR limit, i. e. the potential can generate a maximally entangled state provided that the state momentum is high enough.

The behavior of entanglement generated by a pseudovector potential depends on whether the spatial component of the potential, $\vec{W}$, is perpendicular to the momentum, $\vec{\mathcal{P}}$, or if the time component of the potential, $q$, is null. For the former case, entanglement has maximum values depending on $q$ and vanishes for $q \gg \sqrt{m^2 + \mu^2}$. For the latter case, entanglement is a decreasing function of $\cos \theta$, where $\theta$ is the angle between $\vec{W}$ and $\vec{\mathcal{P}}$, and therefore it vanishes for $\theta = 0$.

The last class of potentials consists of a Hamiltonian containing the tensor, pseudovector, and pseudoscalar potentials. For a null time component of the pseudovector field, $q \sim 0$, if $\vec{W}$ is perpendicular to both the magnetic field and momentum, entanglement has an abrupt change of behavior at $\sin \theta = \sin \theta_c = \mu W / \mathcal{P} B $.
If $W = \mu$, it corresponds to a transition from an entangled state to a maximally entangled state. Also for $\mu W > \mathcal{P} B$, the state is maximally entangled for any $\theta$ depending on the value of $s$ and the sign of $\vec{W}\cdot\vec{\omega}_B$.
If the magnetic field is perpendicular to both $\vec{W}$ and $\vec{\mathcal{P}}$, for specific values of the parameters, entanglement presents two minimum points.

Approaches to density matrices of Dirac bi-spinors often focus on deriving information content of such states when transformations to inertial and non-inertial frames are performed \cite{n030,n031,n032}. In such applications, the dynamics is described by a free Dirac Hamiltonian and are not straightforwardly generalized to include external fields. Although our approach is based on an algebraic \textit{ansatz}, it recovers known results, such as the free particle limit, and it can be generalized to include all kinds of interacting potentials and localization effects. Therefore, our density operator exhibits an effective description of any Dirac dynamics.

Considering that low-energy excitations of a nonrelativistic electrons in the single layer graphene exhibits a massless Weyl spinor structure often related to $2D$ Dirac equation solutions supported by a $\mbox{SU}(2)$ structure, the quantum separability of electron-electron or electron-hole described through such structures can be quantified under different circumstances of interaction.
For instance, when imperfections are present in graphene, Dirac equation with external fields effectively describes de low-energy excitations \cite{n018} and in the framework of trapped ions the Jaynes-Cummings Hamiltonian can be engineered in order to simulate the dynamics of Dirac equation with external fields \cite{n017}. In such systems, entanglement can be quantified through the framework present here and it can be generalized to include localization and dissipation effects, such as decoherence \cite{n021}. A phenomenological picture can be drawn using the derived formulas and the investigation of entanglement and correlation contents in these Dirac-like systems can provide new insights in their applications to quantum information and quantum computing tasks.
Analogously, noticing that the trapped-ions work as a flexible platform to map several suitable effects in relativistic Dirac quantum mechanics (as for instance, in discussing the planar diffusion and the $2D$ scattering of the corresponding bi-spinor structures), the framework can also be considered in quantifying the pertinent quantum correlations related to the trapped-ion physics.

To summarize, given that the Dirac equation is the equation of motion for fermionic fields \cite{n020}, the entanglement content of Dirac particles \cite{n033, n034, n035} may also have a formulation extended to the context of Dirac fields, as in the description of fermion mixing \cite{n036} and neutrino oscillations \cite{n037}.

{\em Acknowledgments - The work of AEB is supported by the Brazilian Agencies FAPESP (grant 15/05903-4) and CNPq (grant 300809/2013-1). The work of VASVB is supported by the Brazilian Agency CNPq (grant 140900/2014-4).
The authors are in debit with Prof. Salomon S. Mizrahi for his useful suggestions and ideas about our work}

\pagebreak

\begin{thebibliography}{99}
\vspace{-.5cm}
\bibitem{n001}
L. Lamata, J. Le\'{o}n, T. Sch\"{a}tz, and E. Solano, Phys. Rev. Lett. {\bf 98}, 253005 (2007).
\bibitem{n002}
A.H. Castro Neto, F. Guinea, N.M.R. Peres, K.S. Novoselov, A.K. Geim, Rev. Modern Phys. 81 (2009) 109.
\bibitem{n003}
R. G. Unanyan, J. Otterbach, M. Fleischhauer, J. Ruseckas, V. Kudriasov and G. Juzeliunas, Phys. Rev. Lett. {\bf 105}, 173603 (2010). 
\bibitem{n004}
G. Weick, C. Woollacott, W. L. Barnes, O. Hess, and E. Mariani, Phys. Rev. Lett. {\bf 110}, 106801 (2013). 
\bibitem{n005}
J. Cayssol, C. R.Physique {\bf 14}, 760 (2013). 
\bibitem{n006}
B. M. Rodr\'{\i}guez-Lara and H. M. Moya-Cessa, Phys. Rev. A {\bf 89}, 015803 (2014). 
\bibitem{n007}
L. J. Garay, J. R. Anglin, J. I. Cirac and P. Zoller, Phys. Rev. Lett. {\bf 85}, 4643 (2000). 
\bibitem{n008}
P. M. Alsing, J. P. Dowling, and G. J. Milburn, Phys. Rev. Lett. {\bf 94}, 220401 (2005). 
\bibitem{n009}
M. I. Katsnelson, K. S. Novoselov, and A. K. Geim, Nature Phys. {\bf 2}, 620 (2006). 
\bibitem{n010}
B. Thaller, \textit{The Dirac Equation} (Springer-Verlag, New York, 1992).
\bibitem{n011}
J. J. Sakurai and J. Napolitano, \textit{Advanced Quantum Mechanics} (Addison-Wesley, San Francisco, 2011).
\bibitem{n012}
S. S. Mizrahi, Phys. Scr. {\bf 2009} 014007 (2009).%
\bibitem{n013}
A. E. Bernardini and S. S. Mizrahi, Phys. Scr. {\bf 89} 075105 (2014).%
\bibitem{n014}
A. Peres and D. R. Terno, Rev. Mod. Phys. {\bf 76}, 93 (2004). 
\bibitem{n015}
A. Chodos, R. L. Jaffe, K. Johnson, C. B. Thorn and V. F. Weisskopf, Phys. Rev. D {\bf 9} 3471 (1974). 
\bibitem{n016}
A. Chodos, R. L. Jaffe, K. Johnson and C. B. Thorn, Phys. Rev. D {\bf 10} 2599 (1974). 
\bibitem{nucl001}
J. N. Ginocchio, Phys. Rev. C {\bf 69}, 034318 (2004).
\bibitem{nucl002}
 G. Mao, Phys. Rev. C {\bf67}, 044318 (2003).
\bibitem{nucl003}
R.J. Furnstahl, J. J. Rusnak; B. D. Serot, Nucl. Phys. A {\bf 632} 607 (1998).
\bibitem{n017}
J. Casanova, J. J. Garc\'{\i}a-Ripoll, R. Gerritsma, C. F. Roos and E. Solano, Phys. Rev. A {\bf 82}, 020101(R) (2010). 
\bibitem{n018}
N. M. R. Peres, F. Guinea, and A. H. Castro Neto, Phys. Rev. B {\bf 73}, 125411 (2006). 
\bibitem{n019}
V. A. S. V. Bittencourt, S. S. Mizrahi and A. E. Bernardini, Annals of Physics {\bf 355}, 35-47 (2015)
\bibitem{n020}
C. Itzykson and J. B. Zuber, {\em Quantum Field Theory}, (Mc Graw-Hill Inc., New York, 2006).
\bibitem{n021}
H. P. Breuer, F. Petruccione, {\em The Theory of Open Quantum Systems} (Oxford University Press, New York, 2002).
\bibitem{n022}
R. Horodecki, P. Horodecki, M. Horodecki and K. Horodecki, Rev. Mod. Phys. {\bf 81}, 865 (2009). 
\bibitem{n023}
T. J. Osbourne, Quantum Information and Computation {\bf 7}, 3 pp. 209-227 (2007).
\bibitem{n024}
W. K. Wootters, Phys. Rev. Lett. {\bf 80}, 2245 (1998).
\bibitem{n025}
L. Henderson and V. Vedral, J. Phys. A: Math. Gen. {\bf 34}, 6899-6905 (2001).
\bibitem{n026}
H. Ollivier and W. H. Zurek, Phys. Rev. Lett. {\bf 88}, 017901 (2001). 
\bibitem{n027}
B. Dakic, V. Vedral and C. Brukner, Phys. Rev. Lett. {\bf 105}, 190502 (2010). 
\bibitem{n028}
 L. Michel and A. S. Wightman, Phys. Rev. {\bf 98}, 1190 (1955). 
\bibitem{n029}
A. S. Wightman, Journal of Mathematical Physics {\bf 42}, 674 (2001).
\bibitem{n030}
I. Fuentes-Schuller and R. B. Mann, Phys. Rev. Lett. {\bf 95} 120404 (2005).
\bibitem{n031}
P. M. Alsing, I. Fuentes-Schuller, R. B. Mann and T. E. Tessier, Phys. Rev. A {\bf 74}, 032326 (2006).
\bibitem{n032}
S. Moradi, Phys. Rev. A {\bf 79}, 064301 (2009).
\bibitem{n033}
P. Caban and J. Rembielinski, Phys. Rev. A {\bf 74}, 042103 (2006).
\bibitem{n034}
K. Fujikawa, C. H. Oh and C. Zhang, Phys. Rev. D {\bf 90}, 025028 (2014).
\bibitem{n035}
N. Friis, R. A. Bertlmann, M. Huber and B. C. Hiesmayr, Phys. Rev. A {\bf 81}, 042114 (2010). 
\bibitem{n036}
M. Blasone and G. Vitiello, Annals of Physics {\bf 255}, 2 (1995). 
\bibitem{n037}
M. Blasone, A. Capolupo and G. Vitiello, Phys. Rev. D {\bf 66}, 025033 (2002).

\end{thebibliography}
\end{document}